\documentclass[journal,twocolumn]{IEEEtran}
\usepackage{amssymb}
\usepackage{amsfonts}
\usepackage{amsmath}
\usepackage{algorithm}
\usepackage{algorithmic}

\usepackage{graphicx,subfigure,amsmath,amssymb,cite}

\usepackage[dvips]{color}
\usepackage{float}
\ifCLASSINFOpdf
\else
\fi
\begin{document}
\title{Two High-performance Schemes of Transmit Antenna Selection for Secure Spatial Modulation }

\author{Feng Shu, Zhengwang Wang, Riqing Chen, Yongpeng Wu, and Jiangzhou Wang,~\IEEEmembership{Fellow,~IEEE}

\thanks{This work was in part supported by the National Natural Science Foundation of China (Grant Nos. 61771244, 61702258, 61472190, 61501238 and 61602245), the China Postdoctoral Science Foundation (2016M591852), the Postdoctoral research funding program of Jiangsu Province(1601257C), the Natural Science Foundation of Jiangsu Province (Grants No. BK20150791), and the open research fund of National Mobile Communications Research Laboratory, Southeast University, China (No.2013D02).}
\thanks{Feng Shu, Zhengwang Wang are with School of Electronic and Optical Engineering, Nanjing University of Science and Technology, Nanjing, 210094, China.}
\thanks{Riqing Chen, and Feng Shu are with the College of Computer and Information Sciences, Fujian Agriculture and Forestry University, Fuzhou 350002, China.}
\thanks{Yongpeng Wu is with the Shanghai Key Laboratory of Navigation and Location-Based Services, Shanghai Jiao Tong University, Minhang 200240}
\thanks{Jiangzhou Wang is with the School of Engineering and Digital Arts, University of Kent, Canterbury CT2 7NT, U.K. E-mail: j.z.wang@kent.ac.uk}
}

\maketitle

\begin{abstract}
In this paper, a secure spatial modulation (SM) system with artificial noise (AN)-aided is investigated. To achieve higher secrecy rate (SR) in such a system, two high-performance schemes of transmit antenna selection (TAS), leakage-based and maximum secrecy rate (Max-SR), are proposed and a generalized Euclidean distance-optimized antenna selection (EDAS) method is designed. From simulation results and analysis, the four TAS schemes have an decreasing order: Max-SR, leakage-based, generalized EDAS, and random (conventional), in terms of SR performance. However, the proposed Max-SR method requires the exhaustive search to achieve the optimal SR performance, thus its complexity is extremely high as the number of antennas tends to medium and large scale. The proposed leakage-based method approaches the Max-SR method with much lower complexity. Thus, it achieves a good balance between complexity and SR performance. In terms of bit error rate (BER), their performances are in an increasing order: random, leakage-based, Max-SR, and generalized EDAS.
\end{abstract}


\IEEEpeerreviewmaketitle

\section{Introduction}

\IEEEPARstart{S}{patial modulation (sm)} is emerging as a promising multiple-input-multiple-output (MIMO) communication technology, which exploits both the index of activated transmit antennas and amplitude phase modulation (APM) signals to carry bits information\cite{Mesleh2008Spatial,Jeganathan2008Spatial}. Due to low complexity of transmitter and receiver, SM technology is applicable to energy-efficient scenarios. However, in such a SM system, it is very possible that  the confidential messages are intercepted by unintend receivers due to the broadcast nature of wireless channel. How to achieve a secure transmission in SM systems is becoming a hot research topic in wireless networks.

Recently, physical layer security in MIMO systems have been widely investigated, which exploit the uniqueness and time-varying characteristics of channel to obtain secure transmission against eavesdropper\cite{Wyner1975,Hu2016Robust,Shu2016Robust,Zhao2016Physical,Wang2012Distributed}. Some literature focuses on the investigation of SM systems for secure transmission \cite{Guan2013Sec,Wang2015Secrecy,Wu2015Secret,Wu2016Transmitter,Aghdam2016Physical,Liu2017Secure}. The authors in \cite{Guan2013Sec} derived the secrecy mutual information of SM with finite alphabet input and proposed a precoding scheme to improve security for SM or space shift keying (SSK) systems in\cite{Aghdam2016Physical}. Both \cite{Wang2015Secrecy} and \cite{Liu2017Secure} compressed artificial noise (AN) onto the null-space of the desired channel to interfere unknown eavesdropper. Via transmitting AN at transmitter\cite{Wang2015Secrecy} or full-duplex receiver\cite{Liu2017Secure} may achieve a high secrecy rate (SR) approaching to the spectral efficiency (SE). In\cite{Wu2015Secret} and \cite{Wu2016Transmitter}, the authors generalized precoding-aided spatial modulation (PSM) to secrecy PSM with constructing time-varying precoder\cite{Wu2015Secret} or optimizing the precoder by jointly minimizing the receive power at eavesdropper while maximizing the receive power at desired user\cite{Wu2016Transmitter}. PSM systems utilize the indices of receive antennas to carry bits information, which reduces the complexity of receiver. However, all transmit antennas are activated in PSM systems, which results in the problems of inter-channel interference (ICI) and inter-antenna synchronization (IAS).

On the other hand, transmit antenna selection (TAS) for improving the performance of SM systems was firstly investigated in\cite{Rajashekar2013Antenna}. The authors of \cite{Rajashekar2013Antenna} proposed two TAS methods: capacity-optimized antenna selection (COAS) and Euclidean distance-optimized antenna selection (EDAS). The EDAS scheme showed a better bit error rate (BER) performance and higher complexity than COAS scheme. Therefore, some literature was intended to reduce the computational and search complexity of the EDAS scheme\cite{Ntontin2013A,Zheng2015Further,Yang2016Transmit,Sun2017Transmit}.

To the best of our knowledge, there is no research work concerning how to design TAS methods for secure SM system. In this paper, we will focus on the aspect research and make our effort to address this problem, our main contributions are summarized as follows:
\begin{enumerate}
 \item To reach the bound of SR, a maximum secrecy rate (Max-SR) method is proposed with exhaustive search. In accordance with simulation results, it can show the best SR performance in a large range of SNR.
 \item To reduce the leakage of the confidential messages power for desired user to eavesdropper, a leakage-based TAS method is proposed for secure SM system. Also, its low complexity version, i.e., sorting-based solution, is presented to achieve the same SR performance. Simulation results show that the proposed leakage-based method provides a SR performance being close to that of the Max-SR method with far lower-complexity.
 \item Finally, as a performance benchmark, we also generalize the conventional EDAS method in non-secure SM system and make it become a secure EDAS for secure SM system. Meanwhile, we analyze and compare the complexity for the three TAS methods.
\end{enumerate}

The remainder is organized as follows. In Section II, we describe secure SM system model and give a definition for its average secrecy rate. Subsequently, we propose two TAS methods: leakage-based and Max-SR, and generalize the conventional EDAS scheme to secure SM system in Section III. In Section IV, numerical simulation results are presented. Finally, we make our conclusions in Section V.

Notation: throughout the paper, matrices, vectors, and scalars are denoted by letters of bold upper case, bold lower case, and lower case, respectively. Sign $(\cdot)^{-1}$,~$(\cdot)^H$ denote inverse, conjugate transpose, respectively. Notation $\mathbb{E}\{\cdot\}$ stands for the expectation operation. Matrices $\textbf{I}_N$ denotes the $N\times N$ identity matrix, and $\mathrm{tr}(\cdot)$ denotes matrix trace.

\section{System model}
\subsection{Secure Spatial Modulation}
Consider a typical secure SM system as shown in Fig.~1. In the  system, there are a transmitter (Alice) equipped with $N$ transmit antennas, a desired user (Bob) with $N_b$ receive antennas, and an eavesdropper (Eve) with $N_e$ receive antennas, respectively. Without loss of generality, it is assumed that $N$ is not a power of two, thus we have to select $N_t={2^{\left\lfloor {\log _2^{{N}}} \right\rfloor }}$ out of $N$ transmit antennas for mapping the bits to the antenna index. Notice that there are a total $Q = \left(\begin{array}{l}N\\{N_t}\end{array} \right)$ patterns, represented as $\boldsymbol{\Omega}=\{\boldsymbol{\Omega}_1,\cdots,\boldsymbol{\Omega}_Q\}$, where $\boldsymbol{\Omega}_k$ denotes the antennas set of the $k$th pattern. After Alice chooses one pattern from pattern set $\boldsymbol{\Omega}$ and shares it with Bob through a low speed forward link, she activates one of $N_t$ transmit antennas to emit $M$-ary APM symbol and uses the index of activated antenna to convey partial bits information. As a result, the SE is ${\log _2}{MN_t}$ bits per channel use (bpcu).
\begin{figure}
 \centering
 \includegraphics[width=0.42\textwidth]{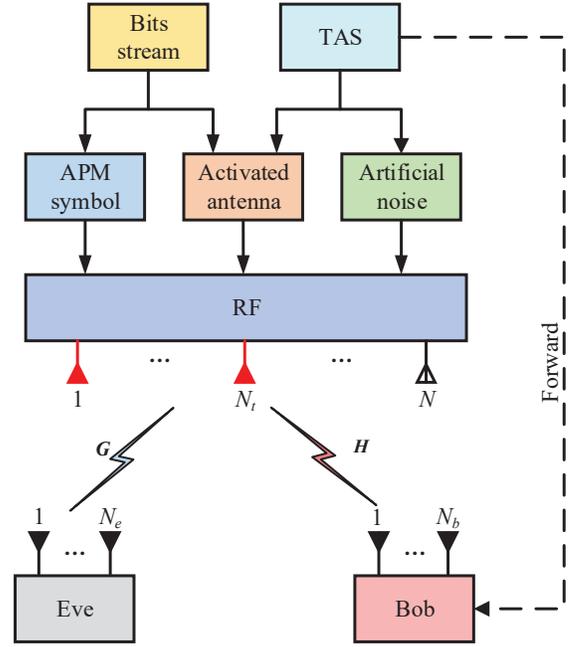}\\
 \caption{System model of secure SM with TAS scheme}\label{system}
\end{figure}

Referring to the secure SM system model in\cite{Wang2015Secrecy}, the transmit baseband signal with the aid of AN  can be expressed as
\begin{align}\label{Sys-mod}
\mathbf{x} = \beta_1\sqrt{P_S}\mathbf{e}_n s_m + \beta_2\sqrt{P_S}\mathbf{P}_{AN}\mathbf{n}
\end{align}
where $\mathbf{e}_n$ is the $n$th column of $\mathbf{I}_{N_t}$ for $n\in\left\{1, 2, \cdots, N_t\right\}$, and $s_m$ denotes the input symbol equiprobably drawn from discrete $M$-ary APM constellation for $n\in\left\{1, 2, \cdots, M\right\}$. We normalize the input symbol power to unit, i.e., $\mathbb{E}{|s_m|^2}=1$. In (\ref{Sys-mod}), matrix $\mathbf{P}_{AN}\in \mathbb{C}^{N_t \times N_t}$ is the AN projection matrix, $\mathbf{n}\sim \mathcal{CN}(0,\mathbf{I}_{N_t})$ is the random AN vector, $P_S$ is the total transmit power, $\beta_1$, and $\beta_2$ are the power allocation factors with $\beta_1^2+\beta_2^2=1$.  Here, it is particularly noted that all selected $N_t$ transmit antennas are used to emit AN but only one of them emits APM symbol. Therefore, such secure SM system alleviates the problems of ICI and IAS compared to conventional secure MIMO systems.

Accordingly, the signals observed at Bob and Eve can be respectively formulated as follows
\begin{equation}
\mathbf{y}_b =\beta_1\sqrt{P_S}\mathbf{H}\mathbf{T}_k \mathbf{e}_n s_m + \beta_2\sqrt{P_S}\mathbf{H}\mathbf{T}_k\mathbf{P}_{AN}\mathbf{n} + \mathbf{n}_b,
\end{equation}
and
\begin{equation}
\mathbf{y}_e=\beta_1\sqrt{P_S}\mathbf{G}\mathbf{T}_k \mathbf{e}_n s_m + \beta_2\sqrt{P_S}\mathbf{G}\mathbf{T}_k\mathbf{P}_{AN}\mathbf{n} + \mathbf{n}_e,
\end{equation}
 where $\mathbf{H}\in\mathbb{C}^{N_b\times N}$ and $\mathbf{G}\in\mathbb{C}^{N_e\times N}$ are the channel state information (CSI) of desired  and eavesdropping channels, respectively. Without loss of generality, we assume $\mathbf{H}$ and $\mathbf{G}$ are frequency-flat Rayleigh fading channels with each element obeying the distribution $\mathcal{CN}(0,1)$ and the CSIs are available at Alice, which might be true for active eavesdropper. $\mathbf{T}_k\in\mathbb{R}^{N\times N_t},k=1,2,...,Q$, is the TAS matrix constituted by the specifically selected $N_t$ columns determined by $\boldsymbol{\Omega}_k$ of $\mathbf{I}_N$. Meanwhile ${\textbf{n}_b}\sim \mathcal{CN}(0,{\sigma _b^2}\mathbf{I}_{N_b})$ and ${\textbf{n}_e}\sim \mathcal{CN}(0,{\sigma _e^2}\mathbf{I}_{N_e})$ denote the complex additive white gaussian noise (AWGN) vectors at Bob and Eve, respectively.

Here, we project the AN lying in the null-space of desired channel, therefore the $\mathbf{P}_{AN}$ is given by
\begin{equation}
\mathbf{P}_{AN}=\frac{1}{\mu} \left[ {{{\mathbf{I}_{{N_t}}} - {\mathbf{H}_k}^H{{\left( {{\mathbf{H}_k}{\mathbf{H}_k}^H} \right)}^{ - 1}}{\mathbf{H}_k}}} \right]
\end{equation}
where $\mu={\parallel {{{\mathbf{I}_{{N_t}}} - {\mathbf{H}_k}^H{{\left( {{\mathbf{H}_k}{\mathbf{H}_k}^H} \right)}^{ - 1}}{\mathbf{H}_k}}} \parallel_\text{F}}$ is the normalized factor of making the identity $\mathrm{tr}(\mathbf{P}_{AN}^H\mathbf{P}_{AN})=1$ holds, and $\mathbf{H}_k=\mathbf{H}\mathbf{T}_k$. It is obvious that $\mathbf{H}_k\mathbf{P}_{AN}=\mathbf{0}$, this means the AN has no impact on Bob.

Assuming Bob has the knowledge of $\mathbf{H}$ and $\mathbf{T}_k$, the maximum-likelihood detector (MLD) is given by\cite{Jeganathan2008Spatial}
\begin{equation}
 [\hat n,\hat m ]= \mathop {\arg \min }\limits_{n \in [1,{N_t}],m \in [1,M]} \parallel {\mathbf{y}_b -  \beta_1\sqrt{P_S}\mathbf{H}{\mathbf{T}_k}{\mathbf{e}_n}{s_m}} \parallel ^2
\end{equation}

\subsection{Average Secrecy Rate and Problem Formulation}
Here, we characterize the security by evaluating average secrecy rate, which is formulated as
\begin{equation}
\bar{R}_s={\mathbb{E}_{\mathbf{H},\mathbf{G}}}{(R_s)}
\end{equation}
with
\begin{equation}
R_s=\left[ {I\left( {\mathbf{x};{\mathbf{y}_b}\left| \mathbf{H},\mathbf{T}_k \right.} \right) - I\left( {\mathbf{x};{\mathbf{y}_e}\left| \mathbf{G},\mathbf{T}_k \right.} \right)} \right]^+
\end{equation}
where $R_s$ denotes the definition of SR in\cite{Wyner1975}. $I\left( {\mathbf{x};{\mathbf{y}_b}\left| \mathbf{H},\mathbf{T}_k \right.} \right)$ and $I\left( {\mathbf{x};{\mathbf{y}_e}\left| \mathbf{G},\mathbf{T}_k \right.} \right)$ are the mutual information over desired and eavesdropping channels, respectively. According to \cite{Wang2015Secrecy}, they are formulated as follows
\begin{equation}\label{mut-inf}
\begin{aligned}
&I\left( {\mathbf{x};{\mathbf{y}_b}\left| \mathbf{H},\mathbf{T}_k \right.} \right)= {\log _2}^{M{N_t}} -\frac{1}{{M{N_t}}}\sum\limits_{n = 1}^{N_t}\sum\limits_{m = 1}^{M}\mathbb{E}_{\mathbf{n}_b}
\\& {{{\log }_2}\sum\limits_{n' = 1}^{N_t}\sum\limits_{m' = 1}^{M} {\exp \left( {\frac{{{{\parallel {{\mathbf{n}_b}} \parallel}^2} - {{\parallel {\boldsymbol{\delta}_{k,n,m}^{n',m'} + {\mathbf{n}_b}} \parallel}^2}}}{{\sigma _b^2}}} \right)} },
\end{aligned}
\end{equation}
and
\begin{equation}
\begin{aligned}
&I\left( {\mathbf{x};{\mathbf{y}_e}\left| \mathbf{G},\mathbf{T}_k \right.} \right)= {\log _2}^{M{N_t}} -\frac{1}{{M{N_t}}}\sum\limits_{n = 1}^{N_t}\sum\limits_{m = 1}^{M} \mathbb{E}_{\mathbf{n}_e^{'}}
\\& {{{\log }_2}\sum\limits_{n' = 1}^{N_t}\sum\limits_{m' = 1}^{M} {\exp \left( {{{{\parallel {{\mathbf{n}_e^{'}}} \parallel}^2} - {{\parallel {\mathbf{W}^{-1/2}\boldsymbol{\alpha}_{k,n,m}^{n',m'} + {\mathbf{n}_e^{'}}} \parallel}^2}}} \right)} }
\end{aligned}
\end{equation}
where $\boldsymbol{\delta}_{n,m}^{n',m'}$, $\boldsymbol{\alpha}_{n,m}^{n',m'}$ and $\mathbf{n}_e^{'}$ are given by
\begin{equation}
\boldsymbol{\delta}_{k,n,m}^{n',m'}=\beta_1\sqrt{P_S}\mathbf{H}\mathbf{T}_k(\mathbf{e}_n s_m-\mathbf{e}_{n'}s_{m'}),
\end{equation}
\begin{equation}
\boldsymbol{\alpha}_{k,n,m}^{n',m'}=\beta_1\sqrt{P_S}\mathbf{G}\mathbf{T}_k(\mathbf{e}_n s_m-\mathbf{e}_{n'}s_{m'}),
\end{equation}
and
\begin{equation}
\mathbf{n}_e^{'}=\mathbf{W}^{-1/2}(\beta_2\sqrt{P_S}\mathbf{G}\mathbf{T}_k\mathbf{P}_{AN}\mathbf{n} + \mathbf{n}_e)
\end{equation}
where $\mathbf{n}_e^{'}\sim \mathcal{CN}(0,\mathbf{I}_{N_e})$ is the AWGN vector, and $\mathbf{W}$ is the covariance matrix of interference plus noise at Eve
\begin{equation}
\mathbf{W}=\beta_2^2P_S\mathbf{G}\mathbf{T}_k\mathbf{P}_{AN} \mathbf{P}_{AN}^H \mathbf{T}_k^H\mathbf{G}^H + \sigma_e^2 \mathbf{I}_{N_e}
\end{equation}

Finally, our objective is to maximize the SR by TAS scheme, which is written as the following optimization problem
\begin{equation}
\begin{aligned}
& \mathop{ \max }~~~~~~~ {R_s}
\\&\text{subject~to}~~\mathbf{T}_k\in \{\mathbf{T}_1,\mathbf{T}_2,...,\mathbf{T}_Q\}
\end{aligned}
\end{equation}
which is an integer or binary optimization problem and is NP-hard. In the next section, we will present three TAS schemes for secure SM system.

\section{Proposed  Transmit Antenna Selection Methods}
In this section, we propose two new TAS schemes: leakage-based and Max-SR, and generalize the conventional EDAS method to the secure SM scenario. Then, we analyze the complexity for the three TAS methods.
\subsection{Proposed Leakage-Based Antenna Selection Method}
\subsubsection{Problem Formulation}
Similar to multi-user MIMO system in\cite{Sadek2007A}, we view the receive power of confidential messages at Eve as the so-called leakage. Thus, the signal-to-leakage-and-noise ratio (SLNR) for the $n$th channel of the $k$th pattern is written as
\begin{equation}
SLNR_n(\mathbf{T}_k) = \frac{{\beta_1^2P_S\parallel{\mathbf{H}{\mathbf{T}_k}{\mathbf{e}_n}\parallel^2}}}{{\beta_1^2P_S\parallel {\mathbf{G}{\mathbf{T}_k}{\mathbf{e}_n}\parallel^2} + N_b\sigma _b^2}}
\end{equation}
It is assumed that all $N_t$ TAs of the selected pattern are activated with equiprobability to transmit confidential messages, then the optimization problem of maximizing SLNR (Max-SLNR) is formulated as
\begin{equation}\label{max-slnr-opt}
\begin{aligned}
& \mathop {\max }~~~~~~ \sum\limits_{n = 1}^{{N_t}} {SLNR_n}(\mathbf{T}_k)
\\&\text{subject~to}~~\mathbf{T}_k\in \{\mathbf{T}_1,\mathbf{T}_2,...,\mathbf{T}_Q\}
\end{aligned}
\end{equation}
with $\mathbf{T}_k$ being the optimization variable. However, the above optimization problem can be solved by exhaustive search with complexity $\mathcal{O}\left(QN_t\right)$ floating-point operations (FLOPs). To lower its complexity, the low-complexity implementation is necessary.

\subsubsection{Low-complexity sorting-based solution}

Considering all transmit antennas are uncorrelated, that is, each transmit antenna results in different SLNR, the optimization problem in (\ref{max-slnr-opt}) is equivalent to choose the largest $N_t$ values  from all $N$ SLNRs. Once we calculate the SLNR values per transmit antenna, we simply arrange them in descending order by sorting method as follow
\begin{equation}\label{sorting}
\underbrace {SLN{R_{{\pi_{\rm{1}}}}} \ge SLN{R_{{\pi_{\rm{2}}}}}\ge\cdots\ge SLN{R_{{\pi_{N_t}}}}}_{{N_t}~{\text{selected~antennas}}}\ge\cdots\ge SLN{R_{{\pi_N}}}
\end{equation}
where $\{{\pi _{1}},{\pi _{2}},...,{\pi _{N}}\}$ is an ordered permutation of $\{1,2,...,N\}$ and $SLNR_{\pi_n}$ is written as
\begin{equation}
SLNR_{\pi_n} = \frac{{\beta_1^2P_S\mathrm{tr}\left( {\mathbf{h}_{\pi_n}} {\mathbf{h}_{\pi_n}}^H\right)}}{{\beta_1^2P_S\mathrm{tr}\left( {\mathbf{g}_{\pi_n}}{\mathbf{g}_{\pi_n}}^H \right) + N_b\sigma _b^2}}
\end{equation}
where ${\mathbf{h}_{\pi_n}}$ and ${\mathbf{g}_{\pi_n}}$ are the $\pi_n$th column of $\mathbf{H}$ and $\mathbf{G}$, respectively.
\subsection{Proposed Max-SR Antenna Selection Method}
In the previous subsection, it is hard for the proposed leakage-based method to maximize SR. In what follows, we will propose the Max-SR method, which aims to maximize SR. For a specific channel realization and the $k$th selected pattern, the transmission rates of desired and eavesdropping channels are respectively bounded as\cite{Rajashekar2013Antenna}
\begin{equation}
\alpha  \le {R_b} \le \alpha  + {\log _2}{N_t}
, ~
\beta  \le {R_e} \le \beta  + {\log _2}{N_t}
\end{equation}
where $\alpha$ and $\beta$ are given by
\begin{equation}
\begin{array}{l}
\alpha  = \frac{1}{{{N_t}}}\sum\limits_{n = 1}^{{N_t}} {\log _2} {\left( 1 + \frac{{{{\beta_1^2P_S\parallel {\mathbf{H}{\mathbf{T}_k}{\mathbf{e}_n}s_m} \parallel}^{\rm{2}}}}}{{{N_b}\sigma _b^2}} \right)}\\
\beta  = \frac{1}{{{N_t}}}\sum\limits_{n = 1}^{{N_t}} {\log _2}\left( {1 + \frac{{{{\beta_1^2P_S\parallel {\mathbf{G}{\mathbf{T}_k}{\mathbf{e}_n}s_m} \parallel}^{\rm{2}}}}}{{\beta_2^2P_S {{\parallel {\mathbf{G}{\mathbf{T}_k}{\mathbf{P}_{AN}}\mathbf{n}} \parallel}^2} + {N_e}\sigma _e^2}}} \right)
\end{array}
\end{equation}
Then, the corresponding SR is bounded as
\begin{equation}
\alpha  - \beta  - {\log _2}{N_t} \le {R_s} \le \alpha  - \beta  + {\log _2}{N_t}
\end{equation}

Notice that $\alpha$ and $\beta$ are the functions of $\mathbf{T}_k$, thus we can carefully select a pattern for maximizing $\alpha-\beta$, which can be formulated as
\begin{equation}\label{Max-SR}
\begin{aligned}
&\max ~~~~\sum\limits_{n = 1}^{{N_t}} {\log _2}{\left( {\frac {1 + \frac{{{{\beta_1^2P_S\parallel {\mathbf{H}{\mathbf{T}_k}{\mathbf{e}_n}s_m} \parallel}^{\rm{2}}}}}{{{N_b}\sigma _b^2}}}
{1 +\frac{{\beta_1^2P_S\parallel {\mathbf{G}{\mathbf{T}_k}{\mathbf{e}_n}s_m} \parallel}^{\rm{2}}} {{\beta_2^2P_S {{\parallel {\mathbf{G}{\mathbf{T}_k}{\mathbf{P}_{AN}}\mathbf{n}} \parallel}^2} + {N_e}{\sigma _e^2} }}} } \right)}
\\&\text{subject~to}~~\mathbf{T}_k\in \{\mathbf{T}_1,\mathbf{T}_2,...,\mathbf{T}_Q\}
\end{aligned}
\end{equation}
In general, the optimal TAS pattern for (\ref{Max-SR}) is usually obtained by an exhaustive search approach due to the fact that the $\mathbf{P}_{AN}$ in (\ref{Max-SR}) is associated with the selected pattern.

\subsection{Generalized Euclidean Distance Antenna Selection Method }
Now, we develop the generalized EDAS method. It is well-known that the mutual information lacks closed-form expression for discrete-input continuous-output memoryless channels (DCMC). Here, we present a lower bound for (\ref{mut-inf})
\begin{equation}\label{lb-rate}
\begin{aligned}
&I\left( {\mathbf{x};{\mathbf{y}_b}\left| \mathbf{H},\mathbf{T}_k\right.} \right)_{LB}= \log _2^{M{N_t}} + {N_b}\left( {1 - \frac{1}{{{{\ln }2}}}} \right) - \frac{1}{{M{N_t}}}\\&~~~~~~~\sum\limits_{n = 1}^{N_t}\sum\limits_{m = 1}^{M} {{{\log }_2}\sum\limits_{n = 1}^{N_t}\sum\limits_{m = 1}^{M} {\exp \left( {\frac{{ -\parallel \boldsymbol{\delta}_{n,m}^{n',m'}\parallel^2}}{{2\sigma _b^2}}} \right)} },
\end{aligned}
\end{equation}
which is similar to that in \cite{Guan2013On}.
Note that the lower bound is approximately tight validated by Monte-Carlo simulation, so we may use it as an effective design metric for TAS scheme.

Notice that the term $\boldsymbol{\delta}_{k,n,m}^{n',m'}$ in (\ref{lb-rate}) has an important impact on the lower bound and the minimum Euclidean distance, i.e., $d_{\text{min}}=\mathop {\min }\limits_{(n,m) \ne (n',m')} \parallel \boldsymbol{\delta}_{k,n,m}^{n',m'} \parallel^2$, is the dominant term in high signal-to-noise ratio (SNR) region\cite{Guan2013On}. In other words, the bigger $d_{\text{min}}$ is expected to yield a higher mutual information. Therefore, we aim to select a TAS pattern of maximizing the minimum Euclidean distance over desired channel or minimizing the minimum Euclidean distance over eavesdropping channel\cite{Aghdam2016Physical}, which are formulated as follows
\begin{equation}\label{max-bob}
{\mathbf{T}_{{k^ * }}} = \mathop {\arg \max }\limits_{\mathbf{T}_k\in \{\mathbf{T}_1,\mathbf{T}_2,...,\mathbf{T}_Q\}} ~~ \mathop {\min }\limits_{(n,m) \ne (n',m')} \parallel \boldsymbol{\delta}_{k,n,m}^{n',m'} \parallel^2
\end{equation}
\begin{equation}\label{min-eve}
{\mathbf{T}_{{k^ * }}} = \mathop {\arg \min }\limits_{\mathbf{T}_k\in \{\mathbf{T}_1,\mathbf{T}_2,...,\mathbf{T}_Q\}} ~~ \mathop {\min }\limits_{(n,m) \ne (n',m')} \parallel \boldsymbol{\alpha}_{k,n,m}^{n',m'} \parallel^2
\end{equation}

Here, we obtain two TAS patterns by (\ref{max-bob}) and (\ref{min-eve}), then we select the one which results in higher $R_s$. Note that some low complexity methods for efficiently solving the problems in (\ref{max-bob}) and (\ref{min-eve}) can be found in\cite{Ntontin2013A,Zheng2015Further,Sun2017Transmit}.
\subsection{Complexity Analysis and Comparison}

Here, we make a complexity comparison concerning the above three methods. It is obvious that the proposed leakage-based method has the lowest complexity of $N(2N_b+2N_e)+\mathcal{O}(N^2)\approx \mathcal{O}(N^2)$ FLOPs, where the term $\mathcal{O}(N^2)$ is the search complexity of sorting operation. For the proposed Max-SR method, the complexity is about $2QN_t[N_t^2(N_b+N_e)+N_t(2N_b^2+N_e)+2N_e+N_b]\approx \mathcal{O}(QN_t^3)$ FLOPs. The complexity of generalized EDAS method is made up of three terms\cite{Sun2017Transmit}: (a) the computational complexity of the upper triangle matrix of minimum Euclidean distance, (b) the search complexity for optimal pattern, and (c) the complexity of computing secrecy rate for two patterns. The complexities of terms (a) and (b) are given by $\mathcal{O}_a=16M ({{N^2} - N})\left( {\frac{{{N_b} + {N_e} - 2}}{3}} \right) + N\left[ {2\left( {{N_b} + {N_e}} \right) - 1} \right]$ and $\mathcal{O}_b=\left( {{N_t}^2 + 3{N_t} + 3} \right){2^{N - {N_t}}} - \frac{{N + 5N + 8}}{2}$ in \cite{Sun2017Transmit}. The complexity of term (c) is as follows ${\rm{2}}{M^{\rm{2}}}{N_t}^2{N_{noise}}[ { 2(N_t+1)({N_b}+{N_e}) + {N_e}^2} ]+2Q[N_t^2(N_e+N_b)+2N_t(N_b^2+N_e^2)]$, where $N_{noise}$ is the number of noise samples for evaluating the secrecy rate. In summary, it is evident that their complexities are in increasing order: leakage-based, Max-SR, and generalized EDAS.

\section{simulation and numerical results}
In this section, numerical simulation results are presented to analyze and compare the performance of three TAS methods from different aspects: average secrecy rate, cumulative density function (CDF) of secrecy rate and bit error rate (BER) of Bob. Simulation  parameters are set as follows: $N_b=N_e=2$, $N=15$, $N_t=8$, and $\sigma_b^2=\sigma_e^2$.  The SNR is defined as $\frac{\beta_1^2P_S}{\sigma_b^2}$, and  Quadrature Phase Shift Keying (QPSK) modulation is employed.
\begin{figure}[h]
 \centering
 \includegraphics[width=0.5\textwidth]{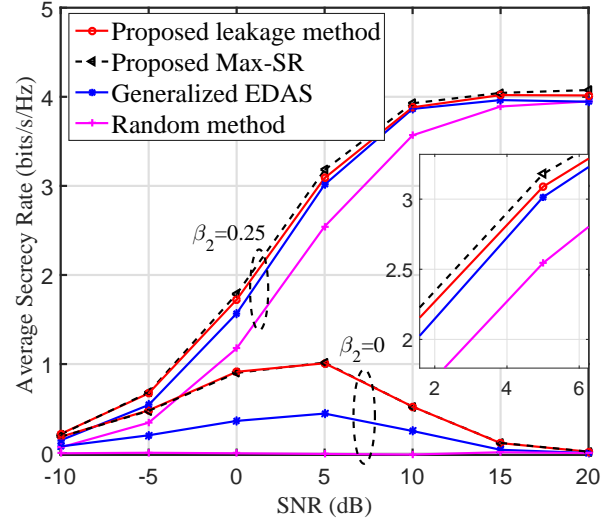}\\
 \caption{Average secrecy rate of four TAS methods for $\beta_2=0.25$, $\beta_2=0$}\label{Fig-SR}
\end{figure}

\begin{figure}[h]
 \centering
 \includegraphics[width=0.5\textwidth]{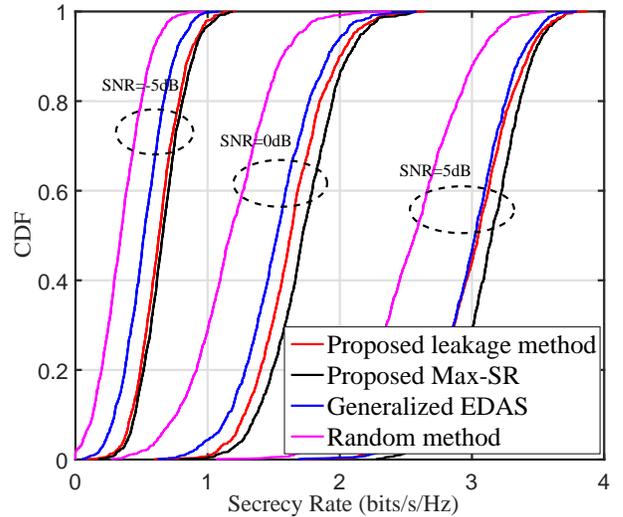}\\
 \caption{CDF curves of secrecy rate for SNR=-5dB, 0dB, 5dB and $\beta_2=0.25$}\label{cdf}
\end{figure}

\begin{figure}[h]
 \centering
 \includegraphics[width=0.5\textwidth]{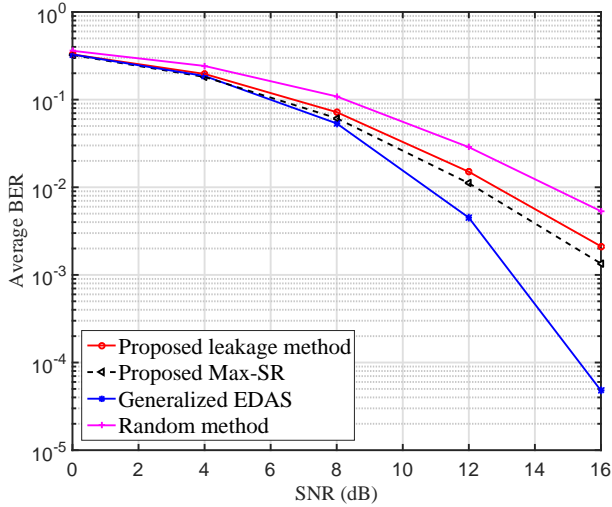}\\
 \caption{Average BER curves of Bob for four TAS methods}\label{ber}
\end{figure}

Fig.~\ref{Fig-SR} demonstrates the average SR for the three TAS methods described in Section-III with $\beta^2_2=0.25$ and $\beta^2_2=0$, where the random TAS method is used for performance reference. For $\beta^2_2=0.25$, the SR performance of the proposed leakage-based method is very close to the Max-SR method and  better than the generalized EDAS method and random method. When we adjust the value of $\beta^2_2$ to zero, we find the SR performance of the proposed leakage-based method is almost identical to that of Max-SR and much better than that of generalized EDAS method. It is also seen that random method fails to achieve positive SR.

Fig.~3 shows the CDF curves of SR for four TAS methods with three different values of SNRs: -5dB, 0dB, and 5dB. It is seen from Fig.~3 that the SR performance of the proposed leakage-based method is between the proposed Max-SR method and the generalized EDAS method. With increasing in the value of SNR, the performance of generalized EDAS method approaches the two proposed methods due to the fact that the minimum Euclidean distance dominates the mutual information term in the high SNR region in accordance with (\ref{lb-rate}). As SNR increases, the CDF curves of SR  moves to the right-hand side. That is, the probability that SR falls in the large value interval grows gradually. However, for all three SNR values, the four TAS methods show the same SR performance trend: Max-SR $\ge$ leakage-based $\ge$ generalized EDAS $\ge$ random.

Fig.~4 illustrates the curves of average BER versus SNR for the three TAS methods described in Section-III with $\beta_2^2=0.25$, where the random method is used for performance reference. From Fig.~4, it is seen that the generalized EDAS method exceeds the remaining three methods in terms of BER performance. This is due to the fact the minimum Euclidean distance makes an important impact on pair-error probability (PEP) in high SNR region. The BER performances of the proposed leakage-based and Max-SR are worse than that of generalized EDAS and better than that of random method.

\section{conclusion}
In this paper, we have made an extensive investigation of TAS methods in secure SM systems. Then, two high-performance TAS schemes: leakage-based and Max-SR, have been proposed to improve the SR performance, and the EDAS method has been generalized to provide a secure transmission. From simulation results and complexity analysis, the proposed Max-SR is the optimal TAS scheme among the three TAS methods in terms of SR performance while the proposed leakage-based method achieves a SR performance near to the proposed Max-SR method with far low complexity. By means of BER performance, the generalized EDAS methods substantially outperforms the remaining two methods due to its main goal to maximize minimum Euclidean distance, which makes a direct improvement on BER performance.

\bibliographystyle{IEEEtran}
\bibliography{IEEEabrv,pls}

\end{document}